\documentclass[journal]{IAENGtran}

\ifCLASSINFOpdf
\usepackage[pdftex]{graphicx}
\DeclareGraphicsExtensions{.png}
\else
\usepackage[dvips]{graphicx}
\DeclareGraphicsExtensions{.eps}
\fi

\begin{document}
	
\title{Multi-Agent Reinforcement Learning with Common Policy for Antenna Tilt Optimization}

\author{Adriano~Mendo, Jose~Outes-Carnero, Yak Ng-Molina and Juan~Ramiro-Moreno
\thanks{Adriano~Mendo, Jose~Outes-Carnero, Yak Ng-Molina and Juan~Ramiro-Moreno are with Cloud and Software Services at Ericsson, 29590 Malaga, Spain (e-mail: \{adriano.mendo, jose.outes, yak.ng.molina, juan.ramiro\}@ericsson.com).}
}

\maketitle

\pagestyle{empty}
\thispagestyle{empty}

\begin{abstract}
%\boldmath
This paper presents a method for optimizing wireless networks by adjusting cell parameters that affect both the performance of the cell being optimized and the surrounding cells. The method uses multiple reinforcement learning agents that share a common policy and take into account information from neighboring cells to determine the state and reward. In order to avoid impairing network performance during the initial stages of learning, agents are pre-trained in an earlier phase of offline learning. During this phase, an initial policy is obtained using feedback from a static network simulator and considering a wide variety of scenarios. Finally, agents can intelligently tune the cell parameters of a test network by suggesting small incremental changes, slowly guiding the network toward an optimal configuration. The agents propose optimal changes using the experience gained with the simulator in the pre-training phase, but they can also continue to learn from current network readings after each change. The results show how the proposed approach significantly improves the performance gains already provided by expert system-based methods when applied to remote antenna tilt optimization. The significant gains of this approach have truly been observed when compared with a similar method in which the state and reward do not incorporate information from neighboring cells.
\end{abstract}

\begin{IAENGkeywords}
	Reinforcement learning, tilt optimization, AI, deep neural network.
\end{IAENGkeywords}

\IAENGpeerreviewmaketitle

\section{Introduction}

\IAENGPARstart{W}{ireless} networks are complex systems, where modification of certain cell parameters may not only affect the performance in that specific cell but also the surrounding cells. Finding an optimal configuration of this kind of parameter might be therefore considered a challenging optimization problem. Examples of these parameters are:
\begin{itemize}\item Remote Electrical Tilt (RET): Defines the angular elevation of a cell antenna and allows for remote modification through a terminal, eliminating the need for physical access to the antenna. Adjusting the RET value may improve the Downlink (DL) Signal to Interference plus Noise Ratio (SINR) in the cell being modified, but at the same time might degrade the SINR of surrounding cells and vice versa.
	\item P0 nominal Physical Uplink Shared Channel (PUSCH): Defines the target power per Physical Resource Block (PRB) that the cell expects in the Uplink (UL), i.e., the communication from the User Equipment (UE) to the Base Station (BS). By increasing it, the UL SINR in the cell under modification increases but, at the same time, the UL SINR in the surrounding cells may decrease, and vice versa.
\end{itemize}

Therefore, there is a clear trade-off between the performance of the modified cell and that of the surrounding cells. This trade-off is not easy to estimate, since it varies on a case-by-case basis. The objective is to optimize the global performance of the network by modifying the RET on a per-cell basis. In computational complexity theory, this kind of problem is considered Non-Polynomial Hard (NP-hard) to resolve.
Numerous scholarly articles and studies have been published on this topic. One of the most common approaches to solving this problem is the use of a control system based on rules defined by an expert. In \cite{Buenestado} a fuzzy rule-based solution is described for RET optimization. With the explosion of artificial intelligence (AI), reinforcement learning (RL) has become a very popular method for solving problems in diverse fields, e.g., autonomous vehicle driving \cite{rhazzaf2021smart}, gaming and puzzle resolution \cite{hukmani2021solving}, and stock trading \cite{li2021enhancing}. Wireless network optimization has also become an attractive field of application, especially in addressing the particular problem of antenna tilt optimization. In \cite{Mismar} a single RL agent for the entire network is proposed, which is no longer valid if new cells are added to the cluster or if some cells become temporarily unavailable. Multi-agent RL (MARL) systems, such as those described by \cite{Guo} and \cite{Koudouridis}, in which each agent acts upon a single cell, perform better in terms of knowledge transfer. In \cite{Balevi} a combination between multi-agent systems and single distributed agents is introduced. However, MARL scenarios are difficult to train because it is necessary to also handle the interaction between individual agents. In \cite{Razavi} a fuzzy system is included as a continuous/discrete converter before an RL agent based on tabular records. Nowadays, there are more efficient ways to handle continuous states, e.g., using neural networks. In some cases, as in \cite{Balevi}, the action of the agent produces the final parameter value to apply. However, RL techniques tend to work better in an incremental fashion, in which the parameter is changed iteratively in small steps, limiting the negative impact of inaccurate reward estimations. A formulation for learning a policy for RET optimization completely offline from real-world network data is successfully applied in \cite{Vannella2020}. However, the performance of off-policy learning is highly sensitive to the quality and variability of data. In \cite{fan2014self} a method is proposed based on fuzzy logic combined with a neural network that considers the impact on neighboring cells. In this study, the observation space is narrow, limiting the proposal to the ideal homogeneous network where it is trained. This paper introduces an alternative RL approach that addresses all the issues mentioned above.

The rest of the paper is organized as follows. Section \ref{rl-overview} describes the basic concepts of RL. Section \ref{approach} reveals the proposed approach for parameter optimization based on RL. Section \ref{methodology} describes the methodology used to generate the results. Results are presented and discussed in Section \ref{results}. Concluding remarks are summarized in Section \ref{conclusion}.

\section{Reinforcement Learning Overview}\label{rl-overview}
RL is a machine learning discipline in which an agent accumulates knowledge about the dynamics of an environment through interactions, with the overall goal of maximizing some notion of cumulative reward. The focus is on finding a balance between the exploration of uncharted territory and the exploitation of current knowledge. Basic RL problems are modeled as a Markov decision process, in which an RL agent (e.g., a network optimizer) interacts with its environment (e.g., the wireless network) in discrete time steps. At each time $t$, the agent receives an observation, which includes the reward $R_t$ and the state $S_t$ \cite{Sutton1998}. Depending on the current state $S_t$, the agent chooses an action $A_t$ from the set of available actions (e.g., parameter changes), which is subsequently applied to the environment. The environment then transitions to a new state $S_{t+1}$ and provides a reward $R_{t+1}$ associated with the transition. The goal of an RL agent is to maximize current and future rewards.
The selection of actions by the agent is modeled as a map called policy. The policy map gives the probability of selecting an action $A_t$ when the environment is in state $S_t$. An example of a policy is epsilon-greedy \cite{Sutton1998}, where the action with the highest expected reward is selected with probability $1-\epsilon$, and a random action is selected with probability $\epsilon$, which is the exploration rate. The concept of discount rate is introduced in many fora to consider the relative impact of future rewards. A discount rate of zero is assumed in this paper, due to the nature of the problem to resolve, in which higher future rewards at the expense of negative or sub-optimal immediate rewards are not acceptable. This research uses an off-policy RL method called Q-learning \cite{Sutton1998}.

\begin{figure}[!t]
	\centering
	\includegraphics[width=3.0in]{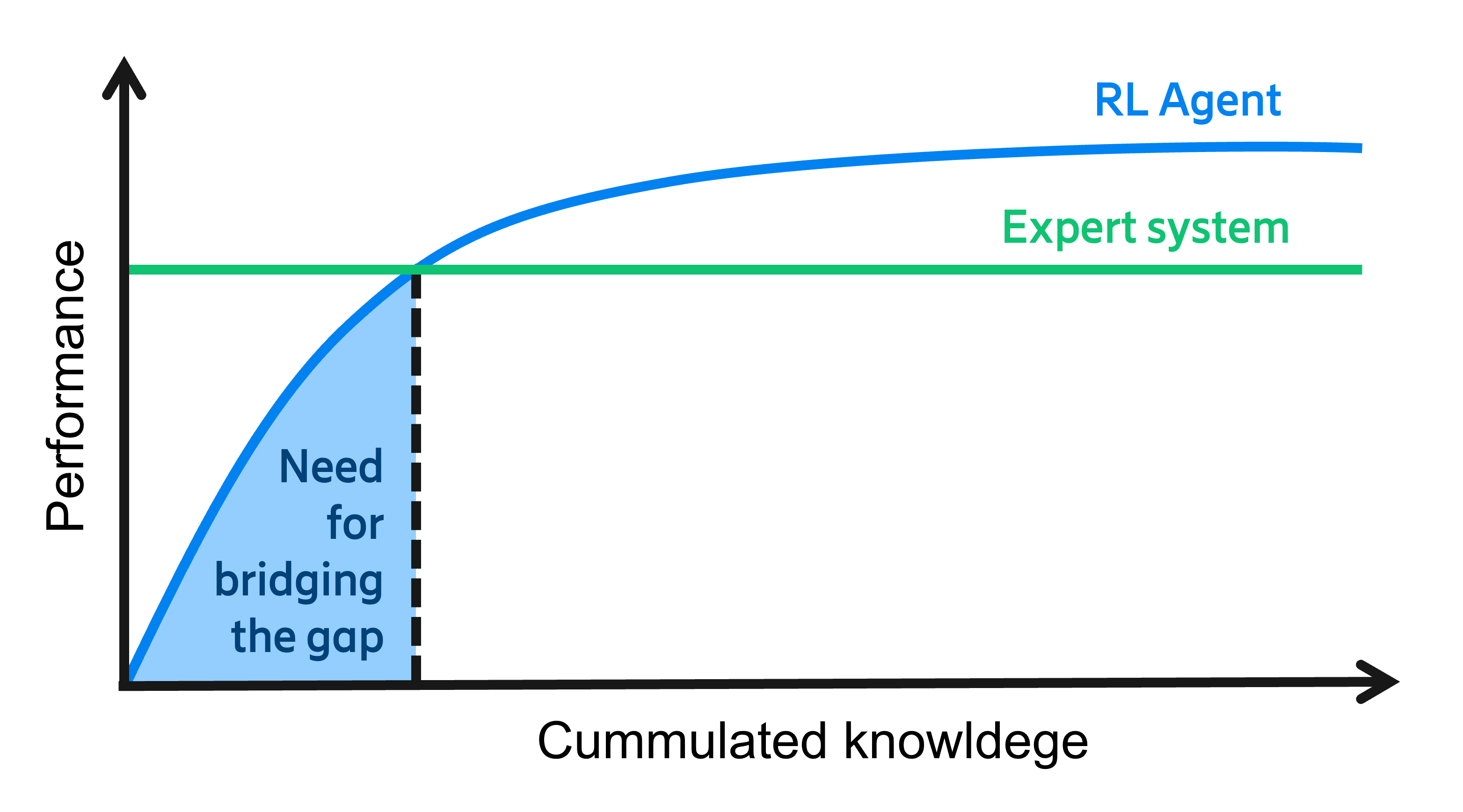}
	\caption{Expected performance evolution of an RL agent and an expert system.}
	\label{fig_rl-vs-expert}
\end{figure}

\section{Proposed Optimization Method Based on Reinforcement Learning}\label{approach}

\subsection{Problem Formulation}
This paper focuses on cellular network optimization problems in which modifying a network parameter in a single cell affects not only the performance of that particular cell but also the performance of the surrounding cells. The study focuses on the case of the RET parameter.

\subsection{Proposed solution}
Policies remain constant in expert systems, which means policies are not improved as expert systems interact with the environment. In contrast, while an RL agent needs to explore the environment to learn, it will eventually surpass the performance of any agent that is defined by an expert, as shown in Fig. \ref{fig_rl-vs-expert}. An offline agent initialization phase can avoid both the initial training phase in the network to optimize, and the corresponding risk of network degradation.

\begin{figure}[!t]
	\centering
	\includegraphics[width=3.4in]{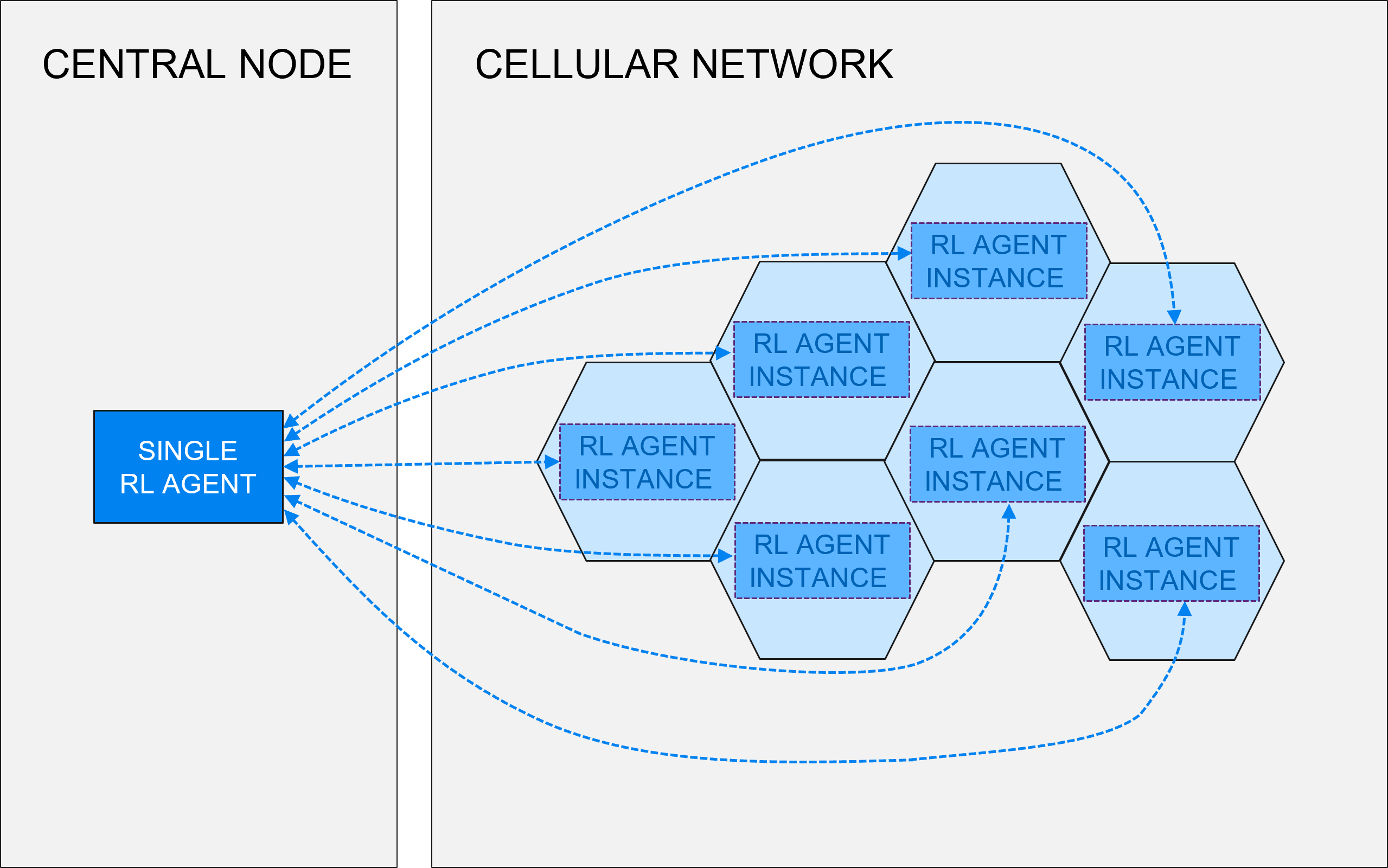}
	\caption{Multiple instances of a unique RL agent.}
	\label{fig_multiple-instances}
\end{figure}

This paper proposes a MARL approach with one agent per cell, where all agents share a common policy. As a consequence, any lessons learned based on rewards obtained from one cell are immediately available in the common policy for the rest of the cells. The agents implement Q-learning with a Deep Neural Network (DNN) and Experience Replay (ER) \cite{lin1992self}. A common policy is obtained by making the agents share the same DNN, which is trained with the experiences collected by all cells. The proposed approach can also be perceived as using multiple instances (one per cell) of a unique RL agent. This conceptual vision is illustrated in Fig. \ref{fig_multiple-instances}, where the single RL agent periodically updates the policy based on the observations collected by the multiple instances. The multiple instances receive the new policy after every update. 

A static network simulator is used to pre-train the DNN during the initialization phase, as proposed in \cite{emr2021ai}. A simulator can provide enough training data variability by just sweeping the required ranges of network configuration parameters. Agents with this same DNN are used to interact with the network to optimize once pre-trained.

The agents steer the cell parameters toward the optimal global solution thanks to suggesting small incremental changes. The use of small incremental changes limits the negative impact caused by errors in reward estimations learned from the training network simulations.

In order to control the interaction between agents, they have visibility of the cell and its surrounding cells in the definition of state and reward. Below, the proposed definitions of state, reward, and action are presented for the particular case of RET optimization of non-active antennas.

\subsubsection{State} In this study, the following configuration parameters and Key Performance Indicators (KPI) have been selected to determine the state of a cell in the RL scheme at every iteration period $t$:
	\begin{itemize}\item Configuration parameters: antenna height, RET, mechanical tilt, carrier frequency, and average distance to the five closest eNodeBs.
	\item KPIs synthesized from Cell Traffic Recordings (CTR) as defined in \cite{Buenestado}: cell overshooting, useless high-level cell overlapping, and bad coverage. The state also includes congestion level, average congestion at the closest cells weighted by their overlapping factor with respect to the studied cell, and an interference indicator, computed as the ratio of traffic in which the RSRP of the second strongest neighbor is higher than a threshold with respect to the best server. The overlapping factor between a first cell and a second cell can be obtained as the periodicity in which both cells are reported simultaneously by the same UE in CTR, provided that the first cell is the serving cell.
	\end{itemize}
\subsubsection{Reward} The reward at instant $t+1$ after a cell parameter update is defined as the relative performance gain after the change, i.e.,
\begin{equation}
 	R_{t+1}=1000 \cdot \frac{RM_{t+1}-RM_{t}}{RM_{t}},
 	\label{eq:reward} 	
\end{equation}
where $RM_{t+1}$ is the reward metric at instant $t+1$ (right after the parameter update), and $RM_{t}$ is the reward metric at instant $t$ (right before the parameter update). This relative definition of the reward improves learning compared to a definition based on absolute values since positive rewards imply a gain, while negative rewards imply a loss. The factor 1000 moves the reward values into a scale that facilitates convergence when using a learning rate of 0.001. The reward metric at instant $t$ is computed as
\begin{equation}
	RM_{t} = 1 + 0.5 \cdot \left(GT_{t} + GT_{t}^{neigh} - CR_{t} - CR_{t}^{neigh}\right),
	\label{eq:rm}	
\end{equation}
where $GT_{t}$ and $CR_{t}$ are the good traffic and the congestion rate at the cell at instant $t$, respectively, and $GT_{t}^{neigh}$ and $CR_{t}^{neigh}$ are the average good traffic and congestion rate measured at instant $t$ at the closest neighboring cells, weighted by their overlapping factor with respect to the studied cell. Good traffic is defined as the ratio of traffic with good coverage and good quality with respect to the total traffic. Good coverage means having reference signal received power (RSRP) over a predefined threshold. Good quality means having DL SINR over a predefined threshold.
\begin{figure}[!t]
	\centering
	\includegraphics[width=3.42in]{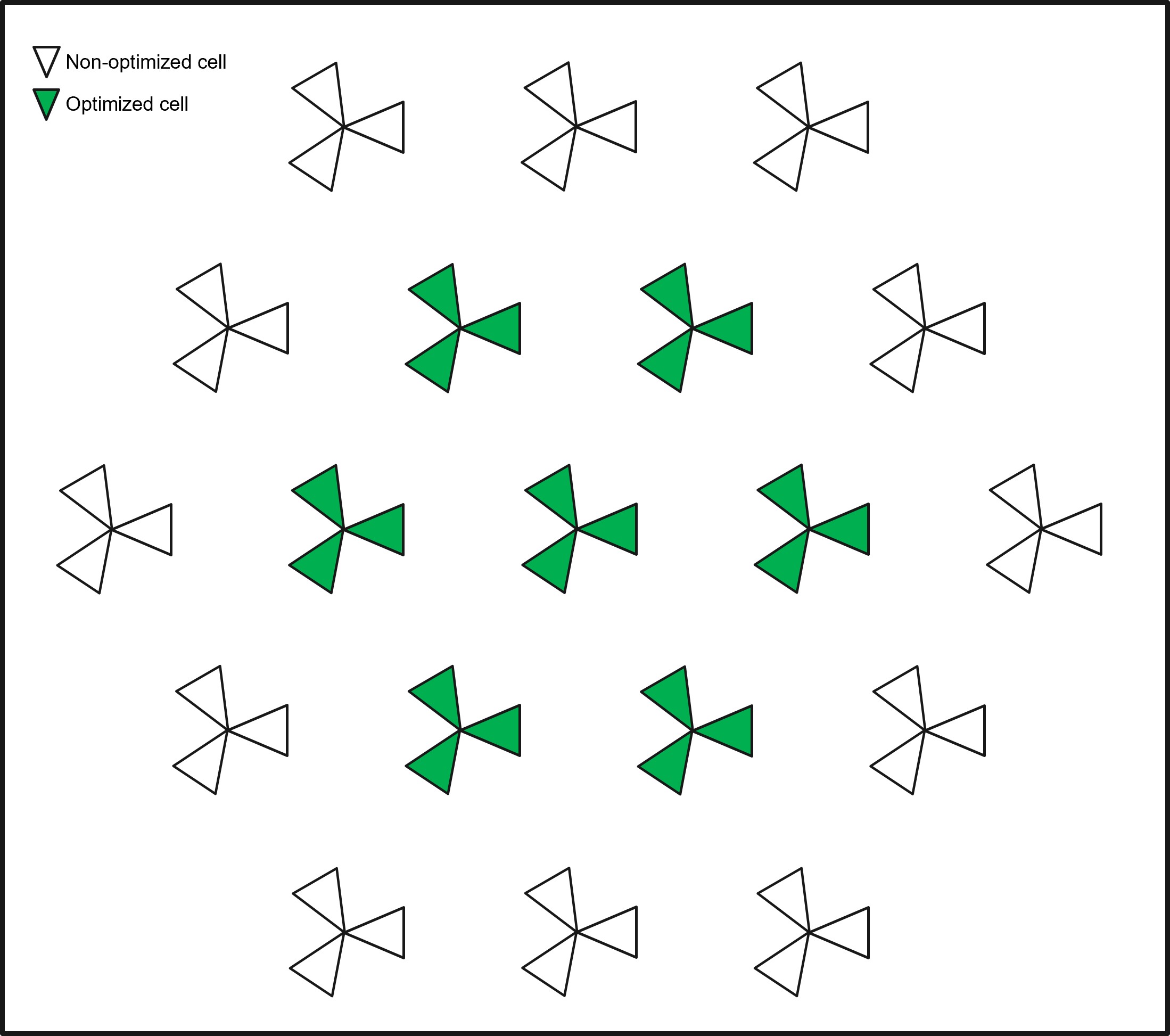}
	\caption{Topology of the scenario used during the training phase.}
	\label{fig_nw-layout-train}
\end{figure}
\subsubsection{Action} This optimization process follows an iterative approach, in which the parameters are not updated with the final value in one single step, but only with a relative increment with respect to the current value. Three actions are possible: keep the current value, increase it by a fixed amount, or decrease it by a fixed amount. A fixed increment of one degree is used in this study. A small increment limits the reward estimation error and the negative impact in case of a wrong decision.

\section{Validation Methodology}\label{methodology}

\subsection{Simulator}
\begin{figure}[!t]
	\centering
	\includegraphics[width=3.46in]{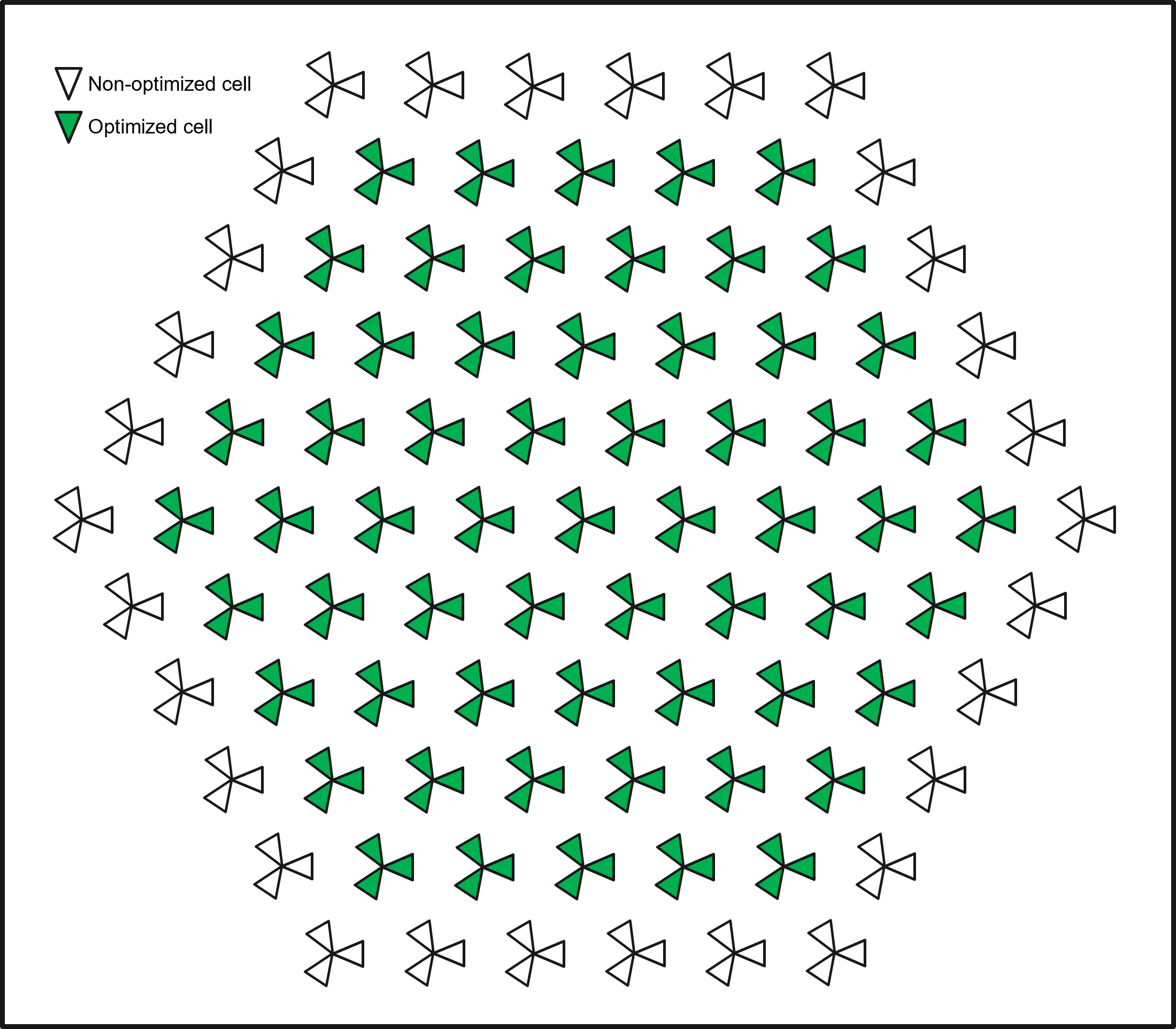}
	\caption{Topology of the scenario used during the test phase.}
	\label{fig_nw-layout-test}
\end{figure}

A proprietary static Monte Carlo simulator has been used to train the agents during the pre-training phase. A large variety of scenarios is considered during this initial phase. In this phase, changes are applied to the training networks, avoiding impacting the test network, which plays the role of the network to optimize. In order to test the performance of the pre-trained agents, the test network has also been simulated, although its topology is different from the one used in the training phase. The regular topology with two cell rings shown in Fig. \ref{fig_nw-layout-train} has been used during the training phase, consisting of 19 sites with three cells per site (i.e., 57 cells). Users are generated following a uniform spatial distribution in the network area. All cells operate on a common single-carrier frequency. The test phase has been carried out in a similar but larger regular scenario, which consists of five rings containing 91 sites (i.e., 273 cells), as displayed in Fig. \ref{fig_nw-layout-test}. Table \ref{tab_sim-params} lists the most relevant network simulation parameters used during the training and test phases. Note that U(A, B) is a discrete uniform distribution in the range [A, B] with step 1. No exploration is carried out during the test phase.

\begin{table}[!t]
	\centering
	\renewcommand{\arraystretch}{1.3}
	\caption{Main Simulation Parameters at the Training and Test Phases.}
	\begin{tabular}{|l||l|}
		\hline
		Parameter&Value\\\hline
		Electrical tilt for optimized cells&U(0, 15) deg\\
		Electrical tilt for non-optimized cells&U(4, 6) deg\\
		Mechanical tilt range&U(0, 4) deg\\
		Antenna height&U(16, 30) m\\
		Inter-site distance& U(1000, 2500) m\\
		Carrier frequencies&0.7, 1.8, 2.1 \& 2.6 GHz\\
		Offered traffic (average per cell)&U(4, 11) Mbps\\
		Bandwidth&20 MHz\\
		Good coverage threshold&-108 dBm\\
		Good quality threshold&3 dB\\
		\hline
	\end{tabular}
	\label{tab_sim-params}
\end{table}

\subsection{Optimization Process}\label{optimization-process}
\begin{figure}[!t]
	\centering
	\includegraphics[width=3.4in]{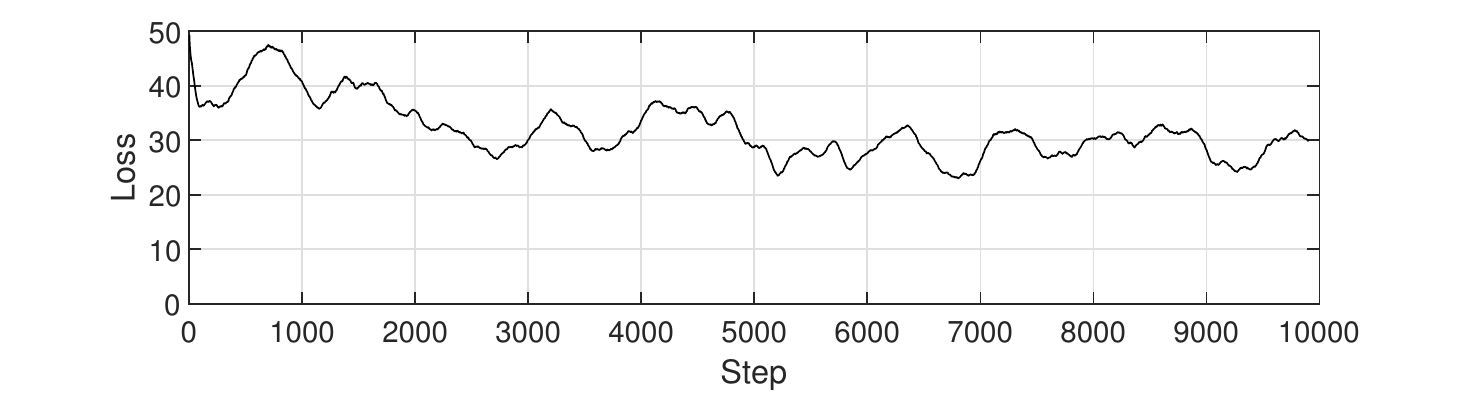}
	\includegraphics[width=3.4in]{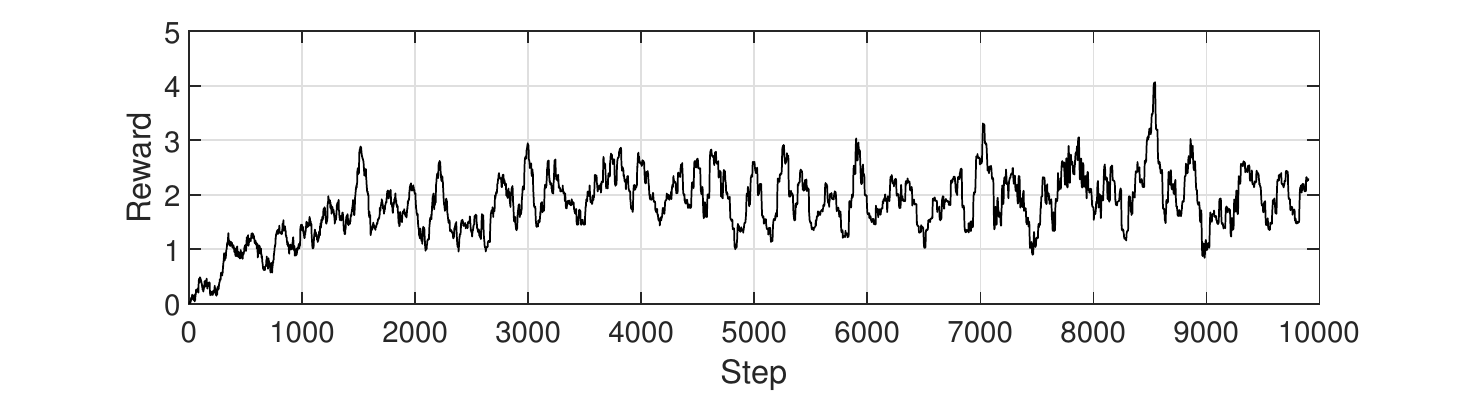}
	\caption{Average loss and reward during the pre-training phase.}
	\label{fig_training-reward}
\end{figure}

 The agents are initially pre-trained offline with 500 independent episodes. Each episode simulates a 20-step optimization campaign, in which the RET of the 21 cells in the inner ring are considered for optimization (see Fig. \ref{fig_nw-layout-train}). The remaining 36 cells in the outer ring keep constant RETs throughout the same episode. These parameters are randomly reset at the beginning of every episode: inter-site distance, mechanical tilt, RET, antenna height, frequency, and offered traffic volume. Mechanical tilt, RET, and antenna height can be different for different cells. During each step of each episode, one of these three possible actions is applied to the RETs of the optimized cells: keep the same value, increase by one degree, or decrease by one degree. Fig. \ref{fig_training-reward} shows the evolution of the loss and the reward during the training phase averaged over a 100-step window so that it is possible to appreciate the long-term trend. The loss function is the mean squared
 reward estimation error.
After the initial learning phase, the agent is tested on 300 additional independent episodes of a larger scenario with a regular topology where 183 out of 273 cells are optimized (see Fig. \ref{fig_nw-layout-test}). To quantify the benefit of including neighbor information, a modified version of the RL agents that excludes the neighbor information from reward and state has also been implemented, pre-trained, and tested on the same 300 episodes. Additionally, the algorithm in \cite{Buenestado}, henceforth referred to as the expert system, has also been implemented to provide results on the same 300 episodes for benchmarking purposes.

\subsection{Evaluation Metric}
The metrics to measure the gain per episode are good traffic improvement, good coverage traffic improvement, good quality traffic improvement, and congestion improvement. The gains per episode are calculated as the relative improvement at the end of an episode compared to the baseline value at the beginning of the same episode.

\section{Simulation Results}\label{results}

\begin{figure}[!t]
	\centering
	\includegraphics[width=3.5in]{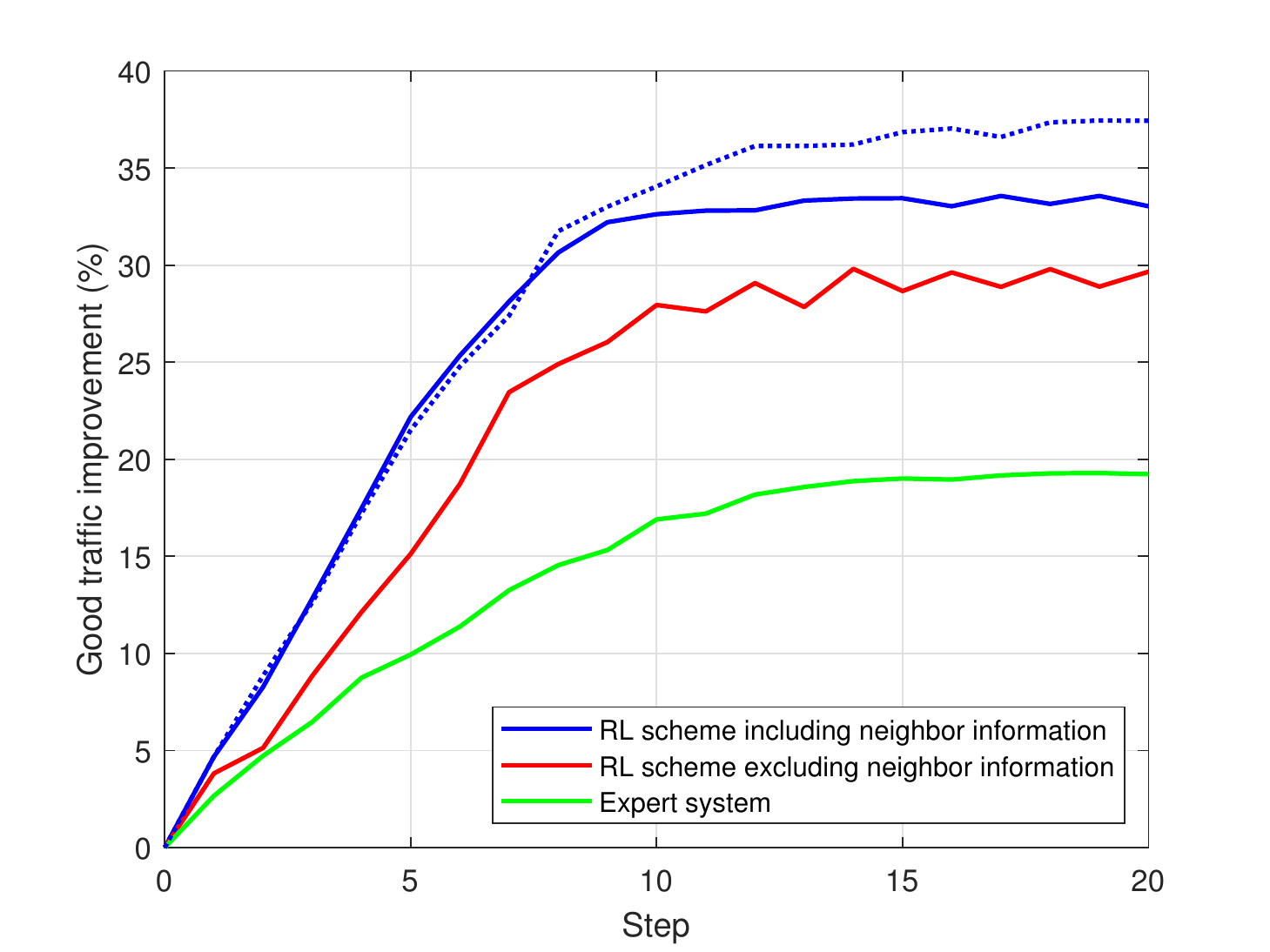}
	\caption{Evolution of average good traffic improvement in a single episode.}
	\label{fig_results-one-episode-good-traffic-gain}
\end{figure}

\begin{figure}[!t]
	\centering
	\includegraphics[width=3.5in]{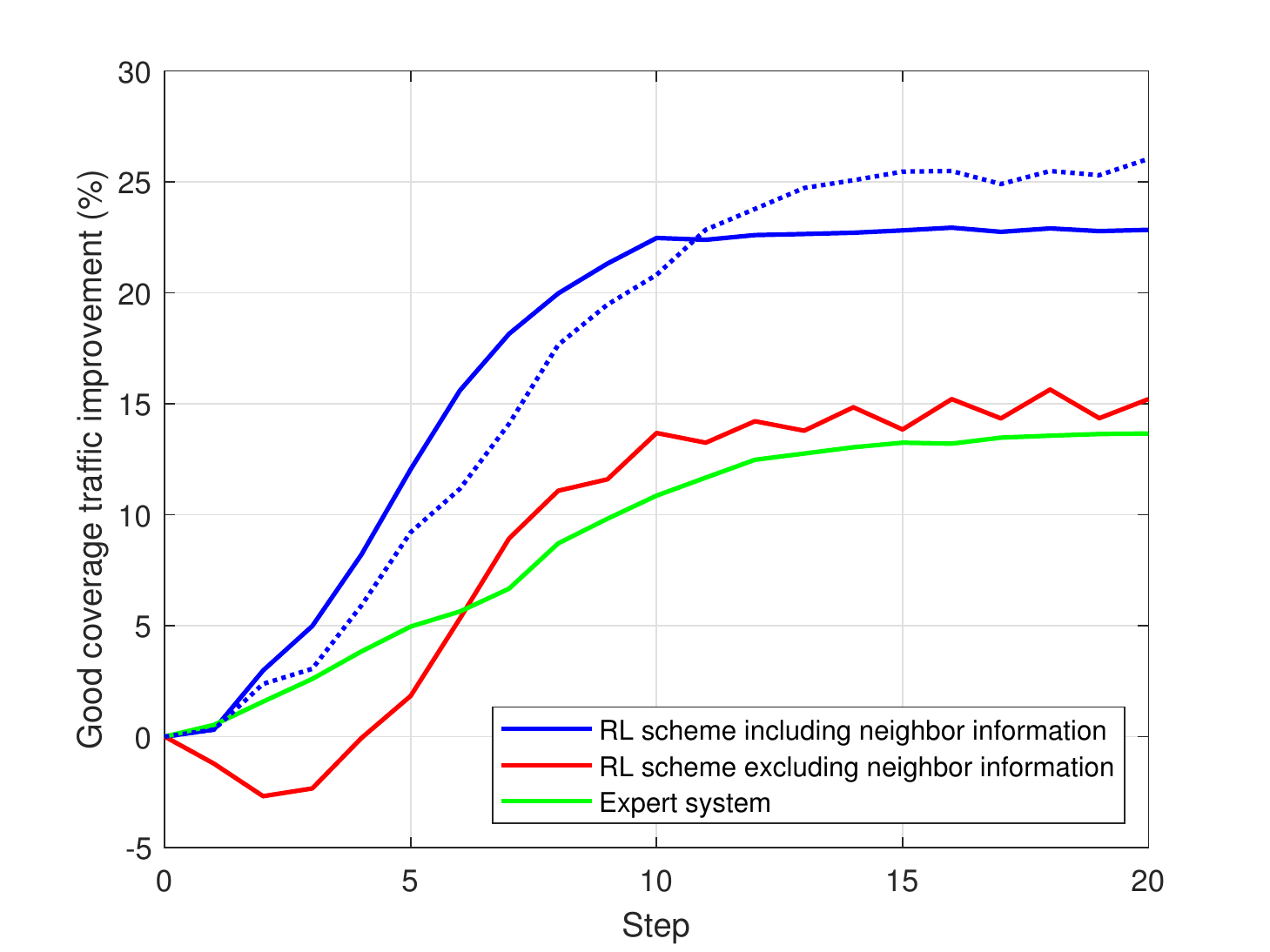}
	\caption{Evolution of average good coverage traffic improvement in a single episode.}
	\label{fig_results-one-episode-good-coverage-traffic-gain}
\end{figure}

\begin{figure}[!t]
	\centering
	\includegraphics[width=3.5in]{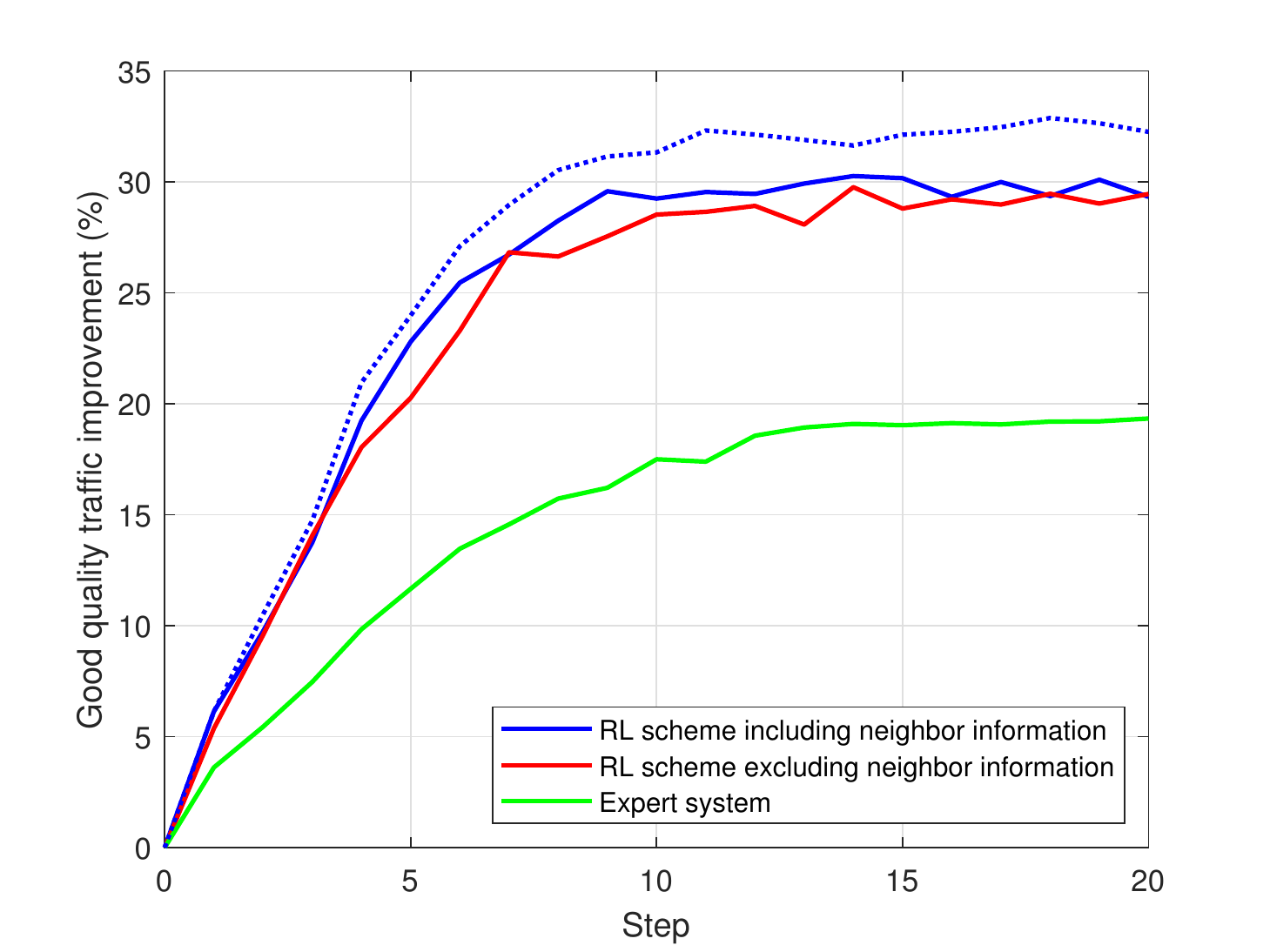}
	\caption{Evolution of average good quality traffic improvement in a single episode.}
	\label{fig_results-one-episode-good-quality-traffic-gain}
\end{figure}

\subsection{Single Episode}
\begin{figure}[!t]
	\centering
	\includegraphics[width=3.5in]{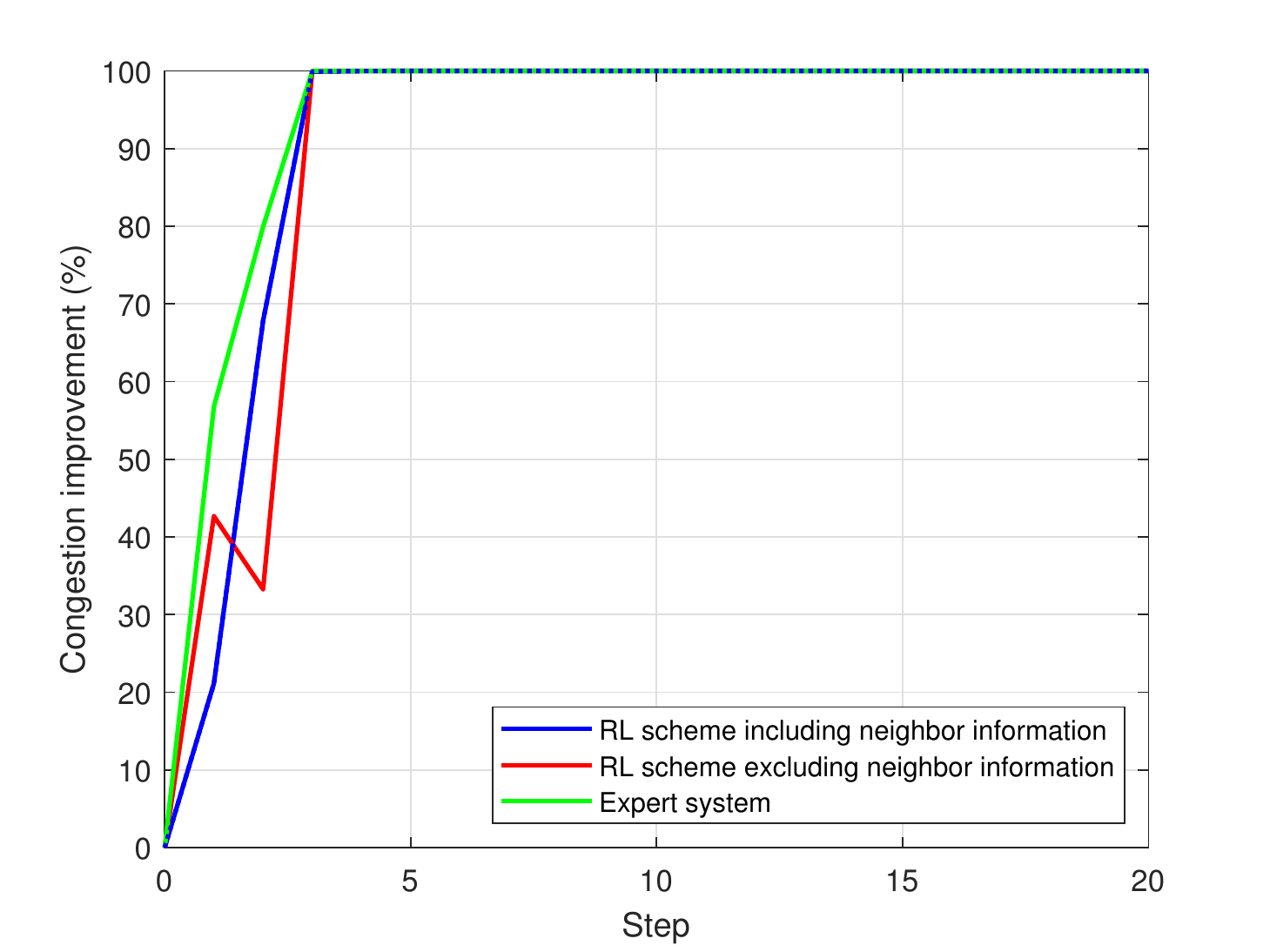}
	\caption{Evolution of average congestion improvement in a single episode.}
	\label{fig_results-one-episode-congestion-gain}
\end{figure}

Fig. \ref{fig_results-one-episode-good-traffic-gain} compares the progress of the average good traffic improvement per step when optimizing the same particular episode with pre-trained RL agents with and without neighbor information, and with the expert system. The solid lines represent the average values using static agents, i.e., agents that do not continue training after the initial pre-training phase. The blue dotted line represents the values obtained with the RL scheme that includes neighbor information and an agent that continues training while optimizing the test network. Step zero represents the baseline configuration. Notice how the RL agents enable greater good traffic improvement from the first step to the last step of the episode, resulting in significantly better traffic improvements than the expert system, particularly when the RL agents utilize neighbor information. This indicates that offline pre-training mitigated the effects of the initial unstable behavior of RL systems. An additional gain is obtained when the agent is configured to continue learning from the network to optimize. This comparison repeats in Figures \ref{fig_results-one-episode-good-coverage-traffic-gain} and \ref{fig_results-one-episode-good-quality-traffic-gain} for the coverage and quality improvements, respectively. A higher quality improvement is also appreciable from the first step when using the RL agents as compared with the expert system. However, the consideration of neighbor information is not so significant as when evaluating the coverage improvement. Fig. \ref{fig_results-one-episode-congestion-gain} shows how the congestion improvement reaches 100\% at the third step with all methods. A congestion improvement of 100\% means that the cells have reduced the congestion to zero. Note that these results have been obtained from a single episode and are presented solely as an illustrative example. Consequently, they are insufficient to draw definitive conclusions. To address this, the experiment has been repeated with a greater number of episodes to ensure statistical significance.

\begin{figure}[!t]
	\centering
	\includegraphics[width=3.5in]{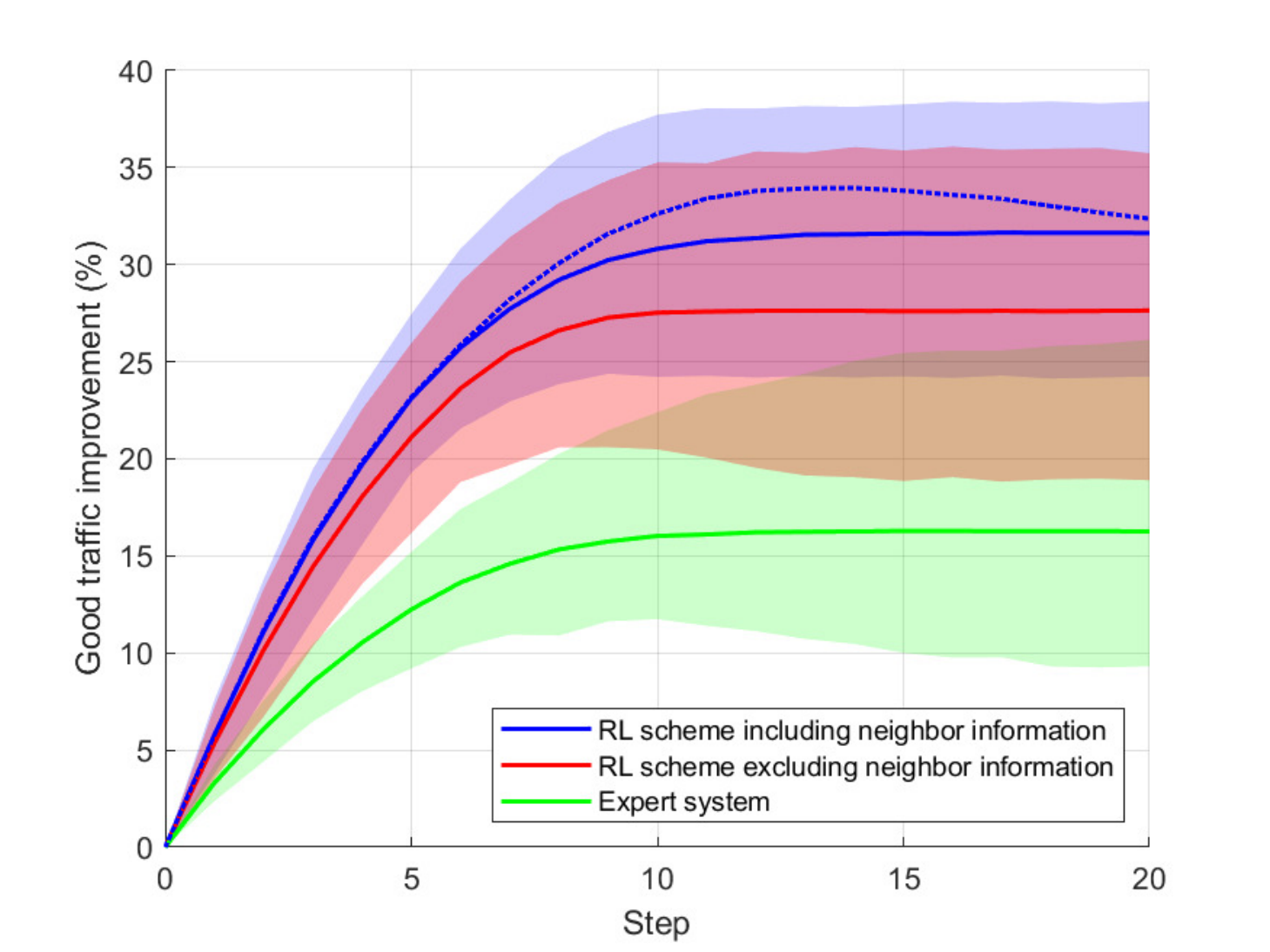}
	\caption{Evolution of good traffic improvement over 300 independent episodes.}
	\label{fig_results-300-episodes-good-traffic-gain}
\end{figure}

\begin{figure}[!t]
	\centering
	\includegraphics[width=3.5in]{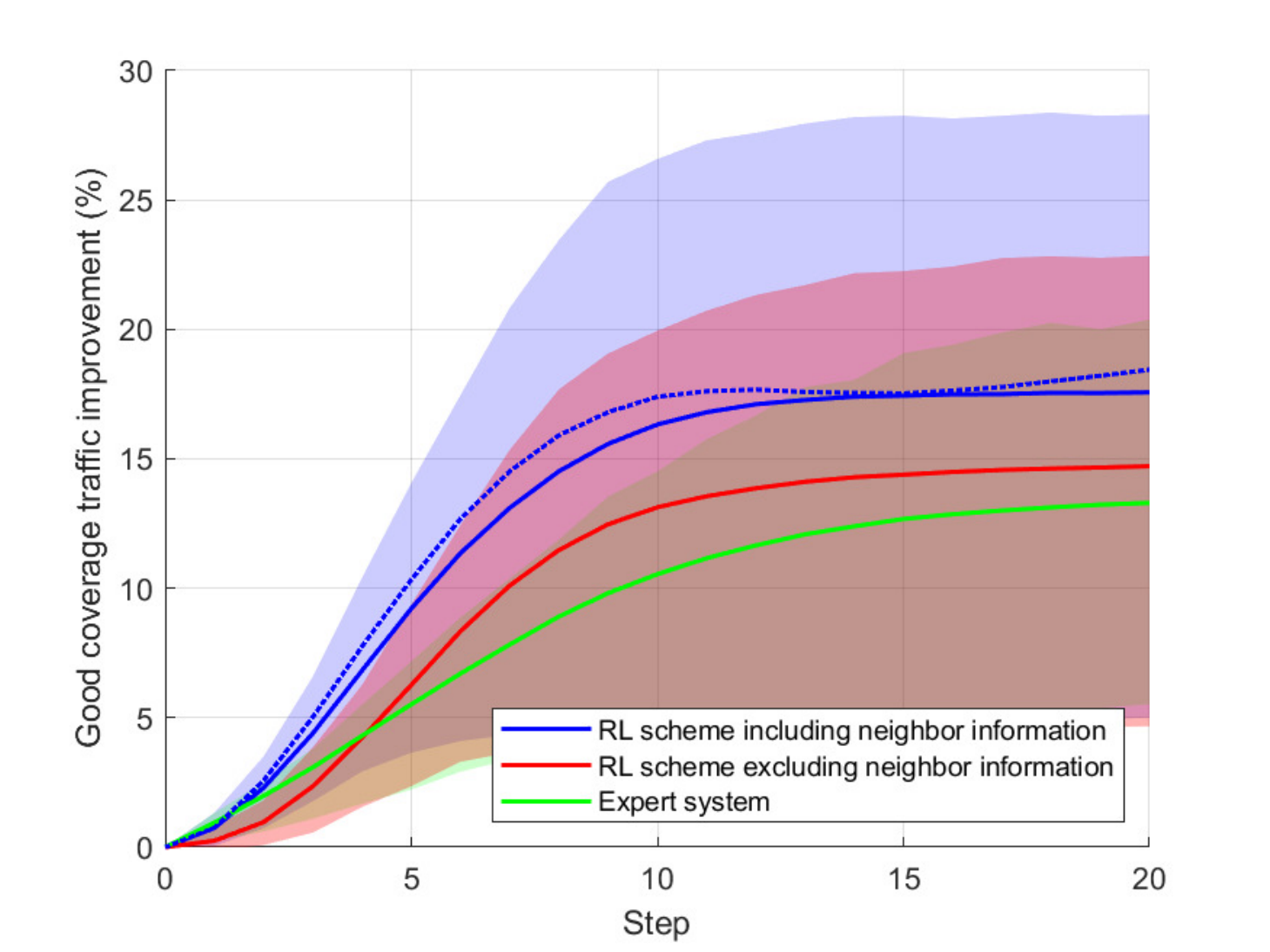}
	\caption{Evolution of good coverage traffic improvement over 300 independent episodes.}
	\label{fig_results-300-episodes-good-coverage-traffic-gain}
\end{figure}

\begin{figure}[!t]
	\centering
	\includegraphics[width=3.5in]{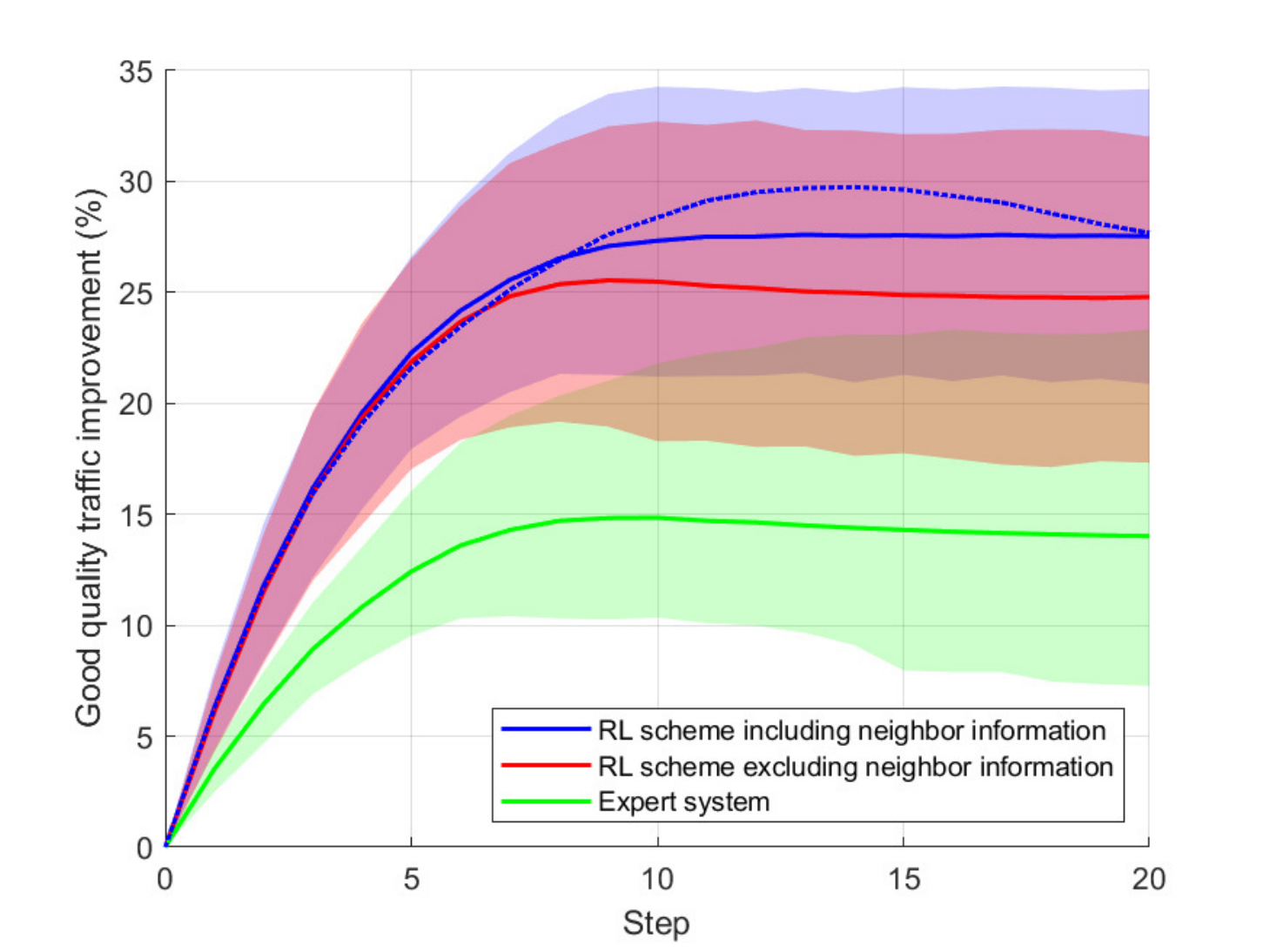}
	\caption{Evolution of good quality traffic improvement over 300 independent episodes.}
	\label{fig_results-300-episodes-good-quality-traffic-gain}
\end{figure}

\begin{table*}[!t]
	\centering
	\renewcommand{\arraystretch}{1.3}
	\caption{Average Final Gains Over 300 Independent Episodes.}
	\begin{tabular}{|l||p{0.8cm}||p{0.8cm}||p{0.8cm}||p{0.8cm}|}
		\hline
		KPI&ES&RLEN&RLIN&RLIN+\\\hline
		Good traffic improvement (\%)&16.3&27.6&31.6&32.4\\
		Good coverage traffic improvement (\%)&13.3&14.7&17.6&18.4\\
		Good quality traffic improvement (\%)&14.0&24.8&27.4&27.7\\
		\hline
	\end{tabular}
	\label{tab_results-three-hundred-episodes}
	\vskip 0.05in
	\raggedright
	\hspace{3.8cm}
	ES: Expert system\\
	\hspace{3.8cm}
	RLEN: RL scheme excluding neighbor information\\
	\hspace{3.8cm}
	RLIN: RL scheme including neighbor information\\
	\hspace{3.8cm}
	RLIN+: RL scheme including neighbor information and continuing to train\\
\end{table*}

\subsection{Multiple Episodes}
The pre-trained RL agents and the expert system have been tested on 300 additional independent episodes, as described in Section \ref{optimization-process}. The results for the multiple episode case are shown in Figures \ref{fig_results-300-episodes-good-traffic-gain} to \ref{fig_results-300-episodes-good-quality-traffic-gain} as the evolution of good traffic improvement, coverage improvement, and quality improvement, respectively, with the number of steps. In this case, the improvements per step are computed as the average values over the 300 episodes, but additional information is graphically provided about the distribution of the gain in terms of the first and third quartiles. The solid lines represent the average values and the shaded areas are the confidence intervals delimited by the first and third quartiles using static agents. The blue dotted line represents the average values obtained with the RL scheme that includes neighbor information and an agent that continues training. The findings are similar to the single-episode case:
\begin{itemize}\item Our RL agent provides significantly higher good traffic improvement than the expert system.
	\item An additional gain is provided by the RL agent when the state and reward definitions include information from neighboring cells, which is especially appreciated in the coverage improvement.
	\item Even higher gain can be achieved if the agent continues to learn from the optimized network.
\end{itemize}

A comparison of the average gains provided by the different studied schemes after the 20 steps of every episode is available in Table \ref{tab_results-three-hundred-episodes}. On average, the proposed RL approach with static agents yields a 94.5\% higher gain than the expert system in terms of good traffic improvement and a 70.0\% higher gain than the same approach when excluding neighbor information.
Another relevant finding is that the proposed RL approach with neighbor information yields a 19.4\% higher coverage gain than the same approach without neighbor information, although the quality gain is only 11.6\% higher. A possible reason for the high quality improvement of the RL scheme that excludes neighbor information is that the quality KPI depends on the SINR, which includes information on interference from neighboring cells.
The RL approach including neighbor information and configured for continuous learning during network optimization achieves the highest performance, resulting in an additional 2.3\% improvement in good traffic compared to using a static agent. As illustrated in the dotted line of Fig. \ref{fig_results-300-episodes-good-traffic-gain}, the optimal performance is not achieved at the last step, but at the 14th step, with a 34.0\% good traffic improvement, i.e., 7.3\% higher gain than using a static agent. This behavior implies that the learning process eventually starts to deteriorate over time. On the one hand, the good coverage traffic is still increasing (dotted line in Fig. \ref{fig_results-300-episodes-good-coverage-traffic-gain}), but on the other hand, the good quality is still decreasing (dotted line in Fig. \ref{fig_results-300-episodes-good-quality-traffic-gain}). By the 15th step, the majority of cells have already achieved optimal RET values, and going forward, the expected behavior of the agent is to keep the configuration unchanged. However, agents may propose oscillations around the optimal RET values. This phase can distort the training of the model, potentially causing it to forget previously acquired knowledge. One potential solution is to exclude training data from cells that have already reached an optimal value. This can be accomplished by detecting consecutive 'keep' actions, oscillating actions, or a decline in reward growth.

\begin{figure}[!t]
	\centering
	\includegraphics[width=3.5in]{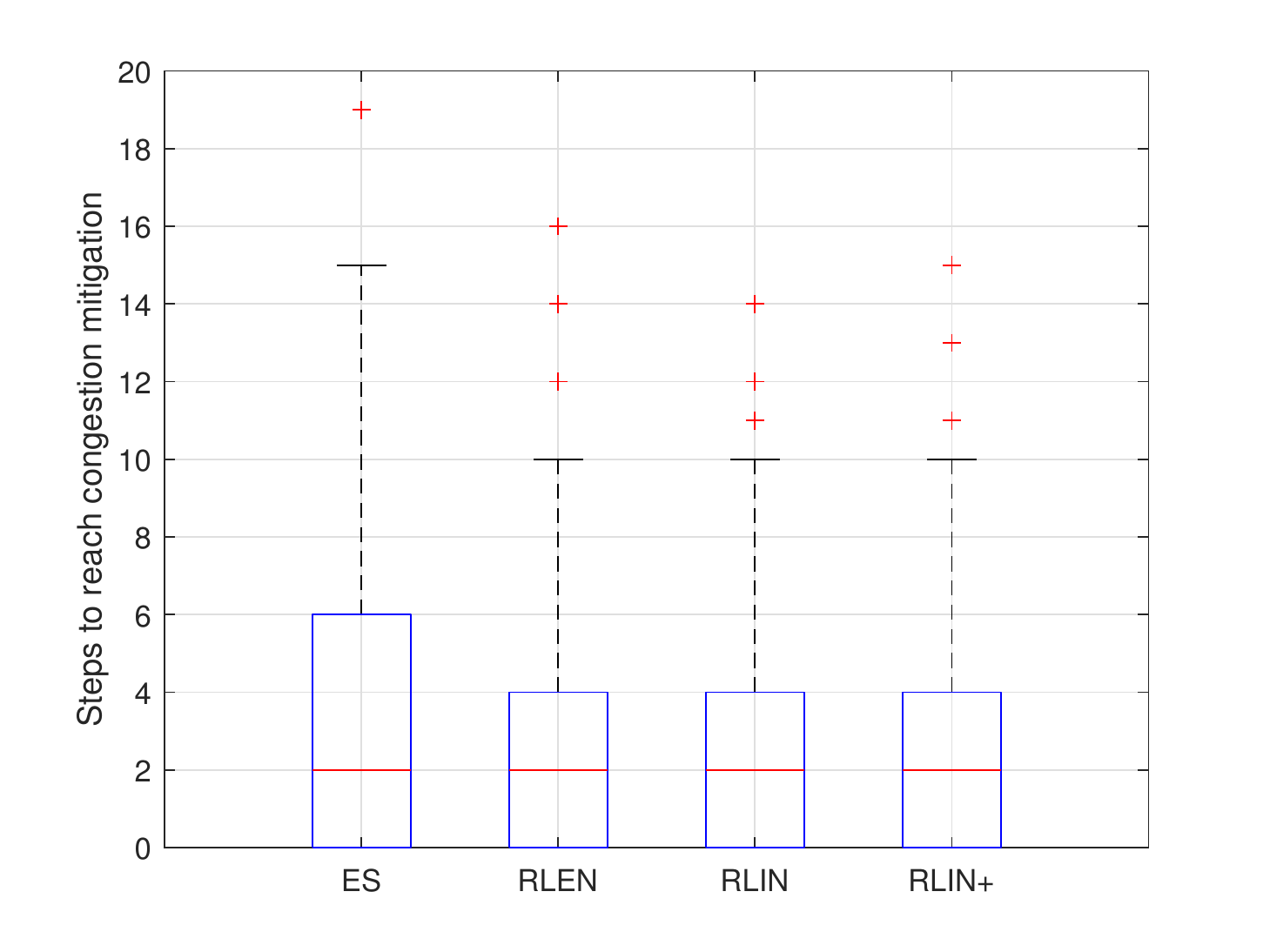}
	\caption{Box plots representing the distribution of the number of steps to reach congestion mitigation over 300 independent episodes.}
	\label{fig_results-300-episodes-steps-to-mitigate-congestion}
\end{figure}

\begin{table}[!t]
	\centering
	\renewcommand{\arraystretch}{1.3}
	\caption{Average Steps to Mitigate Congestion.}
	\begin{tabular}{|l||l||l||l||p{0.7cm}|}
		\hline
		Considered episodes&ES&RLEN&RLIN&RLIN+\\\hline
		All 300 episodes&5.1&3.1&3.0&3.0\\
		Episodes with initial congestion&7.7&4.7&4.6&4.5\\
		\hline
	\end{tabular}
	\label{tab_steps-to-mitigate-congestion}
\end{table}

Congestion improvement consistently reaches 100\% with all methods, which means that all agents can mitigate the congestion. For this reason, rather than looking at the evolution of the average congestion gain, we consider it more insightful to compare the number of steps each method needs to achieve congestion mitigation. Fig. \ref{fig_results-300-episodes-steps-to-mitigate-congestion} displays the distribution of the required steps to reach congestion mitigation for each optimization approach using box plots. The median, represented by the red line, is consistent at two steps across all methods. Table \ref{tab_steps-to-mitigate-congestion} provides the average number of steps needed to achieve congestion mitigation for the different methods. It is noteworthy that the RL-based methods not only deliver significantly improved traffic performance but also exhibit faster congestion mitigation. On average, the expert system requires two more steps compared to the RL methods. However, when focusing on episodes with initial congestion, the RL methods require three fewer steps on average compared to the expert system.

\section{Conclusion}\label{conclusion}
A new approach has been proposed to optimize cell parameters in a wireless network based on MARL, which considers the impact on the cell under study, but also on its neighboring cells. A unique common policy facilitates the immediate sharing of exploration outcomes with the rest of the cells, thereby accelerating the learning phase. It also facilitates knowledge sharing when adding new cells. By employing a simulator for offline pre-training, the network being optimized is protected from potential damage during the initial stages of the learning process, where an untrained agent would exhibit erratic behavior. Simulation results show a significantly higher gain, as compared to an existing expert system in terms of good traffic, which implies coverage increase and interference reduction. Although both methods can mitigate the congestion, the proposed approach does it in fewer steps. Considering neighbor information in the state and reward guarantees significant extra gain, especially in terms of coverage. Finally, after connecting the RL agent to the test network, it can continue learning while optimizing, eventually achieving even higher performance levels. Finding a method to detect the most suitable step to stop learning from a cell is recommended to avoid overfitting, which leads the model to forget previously acquired knowledge.

\ifCLASSOPTIONcaptionsoff
  \newpage
\fi
\bibliographystyle{./BibTeXtran.bst}
\bibliography{MARL-with-Common-Policy_IAENG_2nd_revision_arxiv}
%\newpage
\begin{IAENGbiographynophoto}{Adriano Mendo} received his M.Sc. in Telecommunication Engineering from Malaga University, Spain, in 2004. Since 2004, he has been a Researcher with Optimi Corporation and joined Ericsson in 2010. He has authored a few publications and patents. His current research interests include self-organizing networks, radio resource management, and AI.
\end{IAENGbiographynophoto}
\begin{IAENGbiographynophoto}
	{Jose Outes Carnero} is a Research Specialist at Ericsson in Malaga, Spain. He received his M.Sc. in Telecommunication Engineering from Malaga University, Spain, in 2000. He received his Ph.D. in Electrical and Electronic Engineering from Aalborg University, Denmark, in 2004. His current research interests include network design and optimization, and AI.
\end{IAENGbiographynophoto}
\begin{IAENGbiographynophoto}{Yak Ng Molina} is a Program Manager at Ericsson in Malaga, Spain. He received his M.Sc. in Telecommunication Engineering and his Ph.D. in Electrical and Electronic Engineering from Malaga University, Spain, in 2008 and 2013, respectively. His current research interests include network design and optimization, digital twins, energy management, and AI.
\end{IAENGbiographynophoto}
\begin{IAENGbiographynophoto}{Juan Ramiro Moreno} is heading the Network Design and Optimization Innovation team at Ericsson. He holds a Telecom Engineering degree from Malaga University, a Ph.D. in Electrical and Electronic Engineering from Aalborg University, an MBA from San Telmo Business School, and an Executive Degree in Big Data \& Business Analytics from EOI. He is also an Honorary Professor at Malaga University and co-author of a book on Self-Organizing Networks.	
\end{IAENGbiographynophoto}

\end{document}